\documentclass[useAMS,usenatbib]{mnras}
\usepackage{epsfig}
\usepackage{times}
\usepackage{graphicx}
\usepackage{amsmath}
\usepackage{float}
\def\ra#1#2#3{#1$^{\rm h}$#2$^{\rm m}$#3$^{\rm s}$}
\def\dec#1#2#3{$#1^\circ#2'#3''$}

\title[Endurance of SN~2005ip]{Endurance of SN 2005ip after a decade:
  X-rays, radio, and H$\alpha$ like SN~1988Z require long-lived
  pre-supernova mass loss}

\author[Smith et al.]{Nathan Smith$^1$\thanks{Email:
    nathans@as.arizona.edu}, Charles D.\ Kilpatrick$^1$, Jon C.\
  Mauerhan$^2$, Jennifer E.\ Andrews$^1$, \newauthor Raffaella
  Margutti$^{3,4}$, Wen-Fai Fong$^{1,5}$, Melissa L. Graham$^{2,6}$,
  WeiKang Zheng$^2$, \newauthor Patrick L.\ Kelly$^2$, Alexei V.\
  Filippenko$^2$, Ori D.\ Fox$^7$ \\
  $^1$Steward Observatory, University of Arizona, Tucson, AZ 85721, USA \\
  $^2$Department of  Astronomy, University of California, Berkeley, CA 94720-3411, USA \\
  $^3$Center for Interdisciplinary Exploration and Research in
  Astrophysics (CIERA), and Department of Physics and Astrophysics,
  Northwestern \\ University, Evanston, IL 60208, USA \\
  $^4$Center for Cosmology and Particle Physics, New York University,
  4 Washington Place, New York, NY 10003, USA \\
  $^5$Einstein Fellow \\
  $^6$Department of Astronomy, University of Washington, Box 351580,
   Seattle, WA 98195-1580, USA \\
  $^7$Space Telescope Science Institute, 3700 San Martin Drive,
  Baltimore, MD 21218, USA}

\begin{document}
\date{Accepted 0000, Received 0000, in original form 0000}
\pagerange{\pageref{firstpage}--\pageref{lastpage}} \pubyear{2015}
\def\arcdeg{\degr}
\maketitle
\label{firstpage}

\begin{abstract}
\noindent
Supernova (SN) 2005ip was a Type~IIn event notable for its sustained
strong interaction with circumstellar material (CSM), coronal emission
lines, and infrared (IR) excess, interpreted as shock interaction with
the very dense and clumpy wind of an extreme red supergiant.  We
present a series of late-time spectra of SN~2005ip and a first radio
detection of this SN, plus late-time X-rays, all of which indicate
that its CSM interaction is still strong a decade post-explosion.  We
also present and discuss new spectra of geriatric SNe with continued
CSM interaction: SN~1988Z, SN~1993J, and SN~1998S.  From 3--10 yr
post-explosion, SN~2005ip's H$\alpha$ luminosity and other observed
characteristics were nearly identical to those of the radio-luminous
SN~1988Z, and much more luminous than SNe~1993J and 1998S.  At 10 yr
after explosion, SN~2005ip showed a drop in H$\alpha$ luminosity,
followed by a quick resurgence over several months.  We interpret this
H$\alpha$ variability as ejecta crashing into a dense shell located
$\la0.05$\,pc from the star, which may be the same shell that caused
the IR echo at earlier epochs.  The extreme H$\alpha$ luminosities in
SN~2005ip and SN~1988Z are still dominated by the forward shock at 10
yr post-explosion, whereas SN~1993J and SN~1998S are dominated by the
reverse shock at a similar age. Continuous strong CSM interaction in
SNe~2005ip and 1988Z is indicative of enhanced mass loss for
$\sim10^3$\,yr before core collapse, longer than Ne, O, or Si burning
phases.  Instead, the episodic mass loss must extend back through C
burning and perhaps even part of He burning.

\end{abstract}


%%%%%%%%%%%%%%%%%%%%%%%%%%%%%%%%%%%%%%%%%%%%%%%%%%%%%%%%%%%%%%%%%%%%%%
\section{INTRODUCTION}\label{sec:intro}

Type~IIn supernovae (SNe~IIn) have raised important questions about
the latest phases of evolution in massive stars, because they require
enhanced or episodic mass loss shortly before core collapse that far
exceeds known examples of steady winds (see \citealt{smith14} for a
general review of mass loss and its connection to interacting SNe).
SNe~IIn have relatively narrow lines of hydrogen in their spectra (see
\citealt{avf97} for a review of SN spectral classification) caused by
slow-moving circumstellar material (CSM) that is hit by the blast wave
or illuminated by ultraviolet (UV) radiation from the SN. In some extreme 
cases, as much as 10--25\,M$_{\odot}$ seems to have been ejected in 
just the decade before core collapse
\citep{smith07,woosley07,smith08a,smith10,ofek14}.  In other cases,
the luminosity enhancement is more modest, but the CSM is still dense
enough to correspond to the strongest known winds from extreme and
unstable red supergiants (RSGs; e.g., \citealt{smith09b}).  SNe~IIn
are about 8--9\% of all observed core-collapse SNe
\citep{smith11,li11}, so the physical trigger of the most violent
precursors only operates in a subset of core-collapse events.

A key issue is the timing of this mass loss.  Are these eruptions
synchronized to go off just before the SN owing to instabilities in
the final Ne, O, and Si burning stages of their progenitors
\citep{qs12,sq14,sa14}?  We suspect that pre-SN instability may be
more widespread than indicated by the observed fraction of traditional
SNe~IIn noted above, since these may be only the most extreme
manifestation of a more generic phenomenon \citep{sa14}.  At a less
extreme level, some SNe show Type~IIn signatures for just a day or so
after explosion
\citep{galyam14,smith15,khasov16,nielmala85,quimby07,shivvers15},
which can only be seen if they are discovered early.  At the other
extreme, major mass-loss episodes also seem to precede core collapse
by centuries or millennia.  In the latter case, it could take months
or years for the blast wave to reach the CSM.  This may often go
undetected as follow-up SN observations rarely extend past the first
few months because of sensitivity limits or observers' compulsion to
turn their attention to newer SNe.

There are some observed cases of significantly delayed onset of CSM
interaction.  Most famously, SN~1987A began to collide with its CSM
ring nebula after roughly a decade \citep{sonneborn98,michael98}, and
in this case velocities of the ring nebula indicate that it was
ejected 10--20 thousand years prior \citep{meaburn95,ch00}. This
interaction was relatively modest compared to that of SNe~IIn,
however, and did not lead to a huge increase in luminosity that would
have been observable in a distant galaxy. An interesting case is
SN~2008iy, which did show a large increase in luminosity after a
delay. Initially it had an absolute magnitude of about $-16.5$, similar 
to a normal SN~II-P with no CSM interaction, but then brightened to
about $-19$\,mag after 400 days as the SN caught up
to a massive shell ejected a century earlier \citep{miller10}.
SN~2001em and more recently SN~2014C represent cases in which a
stripped-envelope explosion crashed into a H-rich shell after a year
or more and developed strong, relatively narrow H$\alpha$ emission
\citep{cc06,danny15,raf16}.  Infrared (IR) echoes from CSM dust illuminated 
by the SN itself \citep{wright80} or by ongoing interaction (see, e.g.,
\citealt{fox11}) also provide evidence for massive distant shells.  An
extreme case is SN~2006gy, which --- in addition to its 
$\sim20$\,M$_{\odot}$ shell ejected only 8\,yr before core collapse
\citep{smith10} --- also had another more distant shell of 
10--20\,M$_{\odot}$ ejected $\sim10^3$\,yr earlier
\citep{smith08b,miller10b,fox15}.  Yet another case of a very massive
dust shell seen as an echo is SN~2002hh \citep{barlow05}, although
interestingly, this aging SN does not yet show signs of shock
interaction with this shell even though it probably should have
reached it by now \citep{jen15}.  For SN~2005ip, the SN considered in
this paper, \citet{fox13} interpreted the IR echo properties as
indicating a past major mass-loss outburst that was followed by a less
intense wind.

We therefore have mounting evidence of major mass-loss episodes that
precede core collapse by centuries or millennia, and not just in the
few years beforehand.  If the precursor outbursts of SNe~IIn in the
decade prior to core collapse can be linked to final nuclear burning
phases \citep{qs12,sq14,sa14}, then what can give rise to older mass
ejections?  Wave-driven mass loss during Ne and O burning can only
drive significant mass loss for a few years before core collapse,
according to current models \citep{qs12,sq14}.  Additional
observations are needed to constrain the physical parameters and time
dependence, but there is mounting evidence of enhanced mass loss on
much longer timescales than Ne and O burning.  \citet{woosley16} notes
that the pulsational pair mechanism can produce mass-loss eruptions
that precede core collapse on a variety of timescales, but these come
from very massive stars and should be relatively rare (and the
eruptions might not be punctuated by an energetic SN, as material
falls back to a black hole).  This poses an interesting observational
question: What fraction of SNe (including otherwise normal SNe) have
much more extended dense shells?  For this reason, we have been
monitoring some old, nearby SNe to look for changes in their
late-time interaction.

In this paper we revisit the aging event SN~2005ip, an unusual SN~IIn
located in the host galaxy NGC~2906 at a distance of $\sim$30 Mpc.
\citet{smith09a} presented the optical photometric and spectroscopic
evolution over the first $\sim1000$\,d, while \citet{ori1} discussed
the IR photometric evolution over the same time period.  SN~2005ip was
notable among SNe~IIn for its sustained strong CSM interaction at late
times, its rather unusual high-ionization spectrum with strong coronal
emission lines and a strong blue pseudocontinuum, and its strong IR
emission from dust.  SN~2005ip showed some evidence for the formation
of new dust at these early times \citep{smith09a,ori1}, although
\citet{ori2} concluded that the IR emission was likely produced by a
combination of new dust formation and an IR echo from pre-existing CSM
dust.  \citet{ori2} considered distant dust shells of various radii to
explain the IR echo, and \citet{fox11,fox13} favoured a dust shell
located $\sim0.03$\,pc from the star to explain the IR echo evolution.
This is similar to the radius where we infer late-time CSM interaction
in SN~2005ip, so it is possible that the variability of H$\alpha$
emission that we report here is caused by the ejecta hitting shells
that were seen as an IR echo at earlier times.  This is discussed
below.

Overall, the early-time data suggested a scenario wherein the SN blast
wave was interacting with a very dense and clumpy RSG wind having a
mass-loss rate of more than $2 \times 10^{-4}$\,M$_{\odot}$\,yr$^{-1}$
and likely around $10^{-3}$\,M$_{\odot}$\,yr$^{-1}$
\citep{smith09a}.  This points to an extreme RSG akin to VY~CMa,
rather than a normal RSG like $\alpha$~Ori \citep{smith09b}.

Subsequent papers followed the continuing evolution of SN~2005ip as it
faded during the decade after explosion. By 1800--2400\,d after
discovery, the H$\alpha$ flux had dropped by about a factor of 10
\citep{stritz12}, although this was still much stronger than typical
interacting SNe at a comparable epoch.  \citet{stritz12} also drew
similar conclusions as \citet{smith09a} concerning the evidence for
dust formation and the nature of the progenitor's mass loss, and
\citet{stritz12} mentioned that the X-ray and H$\alpha$ fluxes seem to
provide consistent results. \citet{fox15} also presented a late-time
(day 3024) optical spectrum of SN~2005ip, showing evidence for
continued interaction.  \citet{katsuda14} studied the evolution of
X-ray emission from SN~2005ip during the $\sim2400$\,d after discovery.
They found that the X-ray luminosity was roughly constant until 2009
and dropped by a factor of two by 2012.  They also showed that the
absorbing column density ($N_{\rm H}$) appeared to be steadily declining
during the same time period.  This prompted \citet{katsuda14} to
suggest that the progenitor had ejected a massive CSM shell in the
centuries before explosion, and that after $\sim6$\,yr, the blast wave
was finally emerging from this shell, which they estimated to have a
total mass of $\sim15$\,M$_{\odot}$ swept up by that time.  This
implied that the epoch of CSM interaction was coming to an end in
SN~2005ip.

Here we show that CSM interaction clearly has not yet finished.
The CSM interaction intensity, traced by H$\alpha$ luminosity,
continued to decline very slowly until about mid 2015.  After that
time, however, the H$\alpha$ showed abrupt variations, including a
brightening by late 2015 and early 2016.  The radio and X-ray emission
was also bright in early 2016, as described below.  This indicates
that the blast wave has continued to run into a distant CSM shell or
torus.  We explore possible explanations for the origin of this CSM
and implications for the pre-SN evolution of massive stars.  For
comparison, we also discuss similar continued interaction via
late-time H$\alpha$ emission seen in newly obtained spectra of
SNe~1988Z, 1993J, and 1998S.

\begin{table}  %% 2005ip
\begin{center}\begin{minipage}{3.3in}
    \caption{Late-Time Optical Spectroscopy of SN~2005ip}\scriptsize
\begin{tabular}{@{}lrccc}\hline\hline
  UT Date  &Day$^a$ &Tel./Inst.  & Res.  & $F$(H$\alpha$)$^b$ \\ %\hline       
  (y m d)  & & &$\frac{\lambda}{\Delta \lambda}$ & $10^{-16}$\,erg\,s$^{-1}$\,cm$^{-2}$  \\ \hline
2011 Jan. 13  &1895 &MMT/BC      &4500  &476   \\
2012 Jan. 02  &2249 &MMT/BC      &4500  &259   \\
2012 Apr. 16  &2354 &MMT/BC      &4500  &255   \\
2012 Nov. 24  &2576 &MMT/BC      &4500  &220   \\
2014 May 01  &3099 &Keck/LRIS   &1500  &156   \\
2015 Mar. 23  &3425 &MMT/BC      &4500  &40.8$^c$   \\
2015 Nov. 22  &3668 &LBT/MODS    &1000  &33.9   \\
2015 Dec. 16  &3692 &Keck/DEIMOS &2500  &81.3  \\
2016 Feb. 16  &3755 &MMT/BC      &4500  &113   \\
%2016 Mar. 02  &3770 &Keck/DEIMOS &2500  &315   \\
2016 Mar. 02  &3770 &Keck/DEIMOS &2500  &105   \\ %corrected for NSUM, i.e. *1/3
\hline
\end{tabular}\label{tab:spec}
\end{minipage}\end{center}
$^a$Here and elsewhere, we note ``Day'' as the day from discovery.  For SN~2005ip and SN~1988Z this is an unknown amount after explosion, whereas for SN~1993J and SN~1998S the discovery is likely to have been within a day or so of explosion. \\
$^b$Assumed uncertainty is $\pm25$\%, owing mainly to the systematic uncertainty of
the placement of the standard star within the slit aperture (which is, however, difficult to quantify). Measurement uncertanty due to noise in the spectra is much lower.   \\
$^c$There were some thin clouds that came in during subsequent
exposures (not included in this estimate), so we increased the
uncertainty for this observation when plotted in Figure~\ref{fig:spectra}.
\end{table}

%%%% 88Z
\begin{table}
\begin{center}\begin{minipage}{3.3in}
    \caption{Late-Time Optical Spectroscopy of SN~1988Z}\scriptsize
\begin{tabular}{@{}lrccc}\hline\hline
  UT Date  &Day &Tel./Inst.  & Res.  & $F$(H$\alpha$)$^a$ \\ %\hline       
  (y m d)  & & &$\frac{\lambda}{\Delta \lambda}$ & $10^{-16}$\,erg\,s$^{-1}$\,cm$^{-2}$  \\ \hline
1994 Jun. 10  &2006 &Keck/LRIS   &1500  &67.8   \\
%1995 Nov. 28  &2542 &Keck/LRIS   &1500  &134$^b$   \\ %not corrected
1995 Nov. 28  &2542 &Keck/LRIS   &1500  &56$^b$   \\ % corrected for Hii region
2003 Jan. 06  &5138 &Keck/LRIS   &1500  &21.5   \\
2007 Feb. 16  &6640 &Keck/DEIMOS &2500  &19.6   \\
2010 Feb. 15  &7735 &Keck/DEIMOS &2500  &12.7   \\
2012 May 17  &8557 &Keck/LRIS   &1500  &10.5   \\
\hline
\end{tabular}\label{tab:88z}
\end{minipage}\end{center}
$^a$Assumed uncertainty is $\pm25$\%, owing mainly to the systematic uncertainty of
the placement of the standard star within the slit aperture (which is, however, difficult to quantify). Measurement uncertanty due to noise in the spectra is much lower.\\
$^b$We increase the uncertainty on this point because of possible
calibration issues.  We normalized to the flux of [S~{\sc ii}]
emission lines from underlying H~{\sc ii} regions in the same
aperture at other epochs.
\end{table}

%%%% 93J
\begin{table}
\begin{center}\begin{minipage}{3.3in}
    \caption{Late-Time Optical Spectroscopy of SN~1993J}\scriptsize
\begin{tabular}{@{}lrccc}\hline\hline
  UT Date  &Day &Tel./Inst.  & Res.  & $F$(H$\alpha$)$^a$ \\ %\hline       
  (y m d)  & & &$\frac{\lambda}{\Delta \lambda}$ & $10^{-16}$\,erg\,s$^{-1}$\,cm$^{-2}$  \\ \hline
2004 Nov. 14  &4249 &Keck/LRIS   &1500  &73    \\
2010 Feb. 06  &6159 &Keck/LRIS   &1500  &47    \\
2010 Feb. 15  &6168 &Keck/DEIMOS &2500  &13    \\
2012 Apr. 29  &6972 &Keck/LRIS   &1500  &35    \\
2012 May 17  &6990 &Keck/LRIS   &1500  &43    \\
2013 Feb. 17  &7266 &Keck/DEIMOS &2500  &400     \\ %% check this - NSUM?
2015 Feb. 20  &7999 &Keck/LRIS   &1500  &14   \\
2016 Feb. 10  &8353 &Keck/LRIS   &1500  &82   \\
\hline
\end{tabular}\label{tab:93j}
\end{minipage}\end{center}
$^a$Assumed uncertainty is $\pm30$\%, owing mainly to the systematic
uncertainty of the placement of the standard star within the slit
aperture (which is, however, difficult to quantify), plus some
uncertainty produced by blending with an adjacent oxygen line. Measurement
uncertanty due to noise in the spectra is lower.  \\
\end{table}

\begin{table}
\begin{center}\begin{minipage}{3.3in}
    \caption{SN 2005ip on 2016 Feb. 4; VLA}\scriptsize
\begin{tabular}{@{}ccc}\hline\hline
Frequency & Flux Density & Uncertainty \\ %\hline       
(GHz) & (mJy) & (mJy) \\   \hline
5.0 & 1.18 & 0.04 \\
7.4 & 0.92 & 0.04 \\
8.5 & 0.86 & 0.03 \\
11.0 & 0.77 & 0.03 \\
\hline
\end{tabular}\label{tab:radio}
\end{minipage}\end{center}
\end{table}

%%%%nxrays

\begin{figure*}
\includegraphics[width=7.5in]{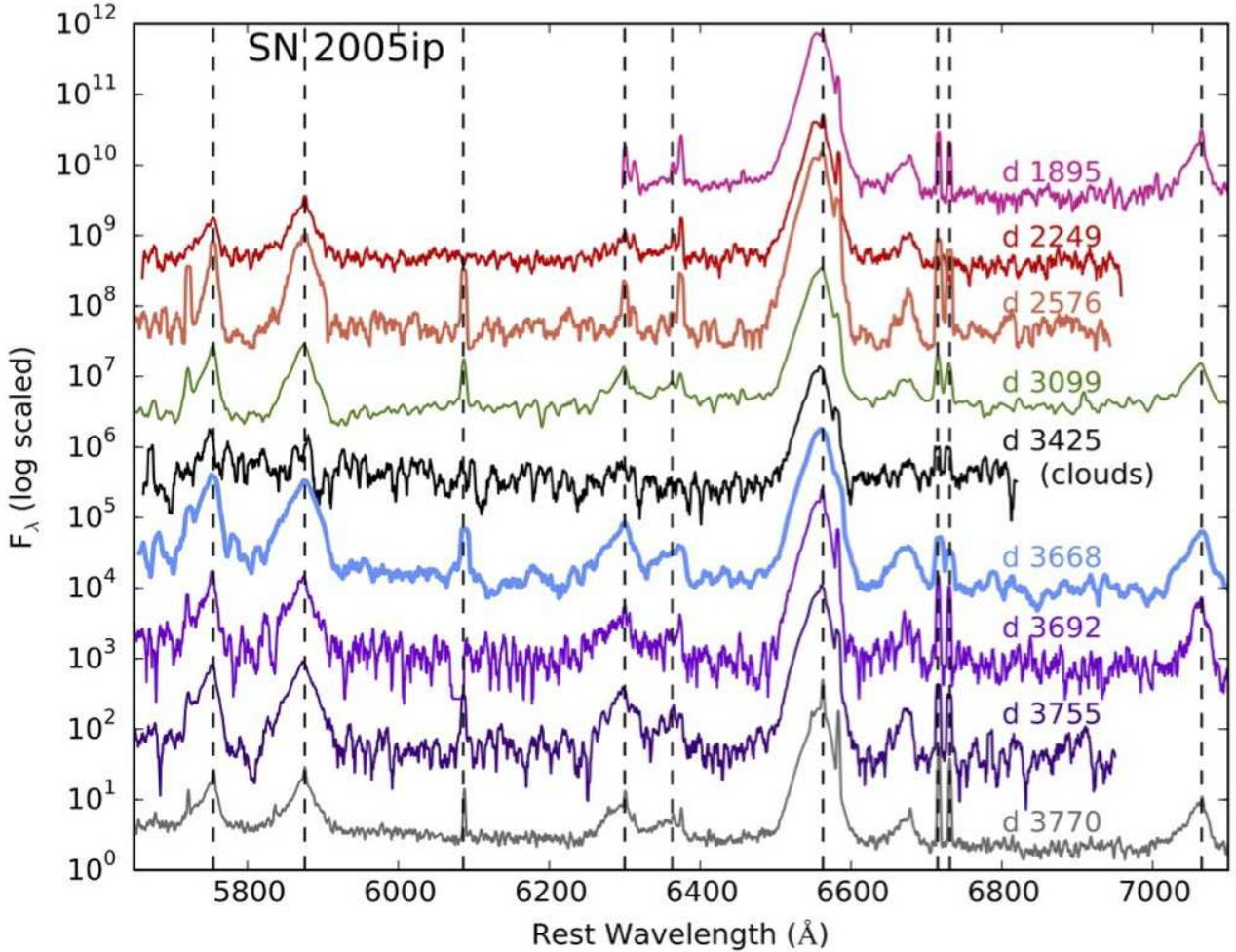}
\caption{Late-time spectra of SN~2005ip (see Table~\ref{tab:spec})
  normalized to the red continuum level. The spectra plotted in thin
  red-orange-green colours represent the appearance of the spectrum in
  its late-time decline.  The black and light-blue spectra are when
  the H$\alpha$ flux reached its minimum in 2015, while the
  purple/blue/grey colours document the resurgence of CSM interaction.
  See Figure~\ref{fig:spectra88z} for line identifications.}
\label{fig:spectra}
\end{figure*}

\begin{figure*}
\includegraphics[width=4.9in]{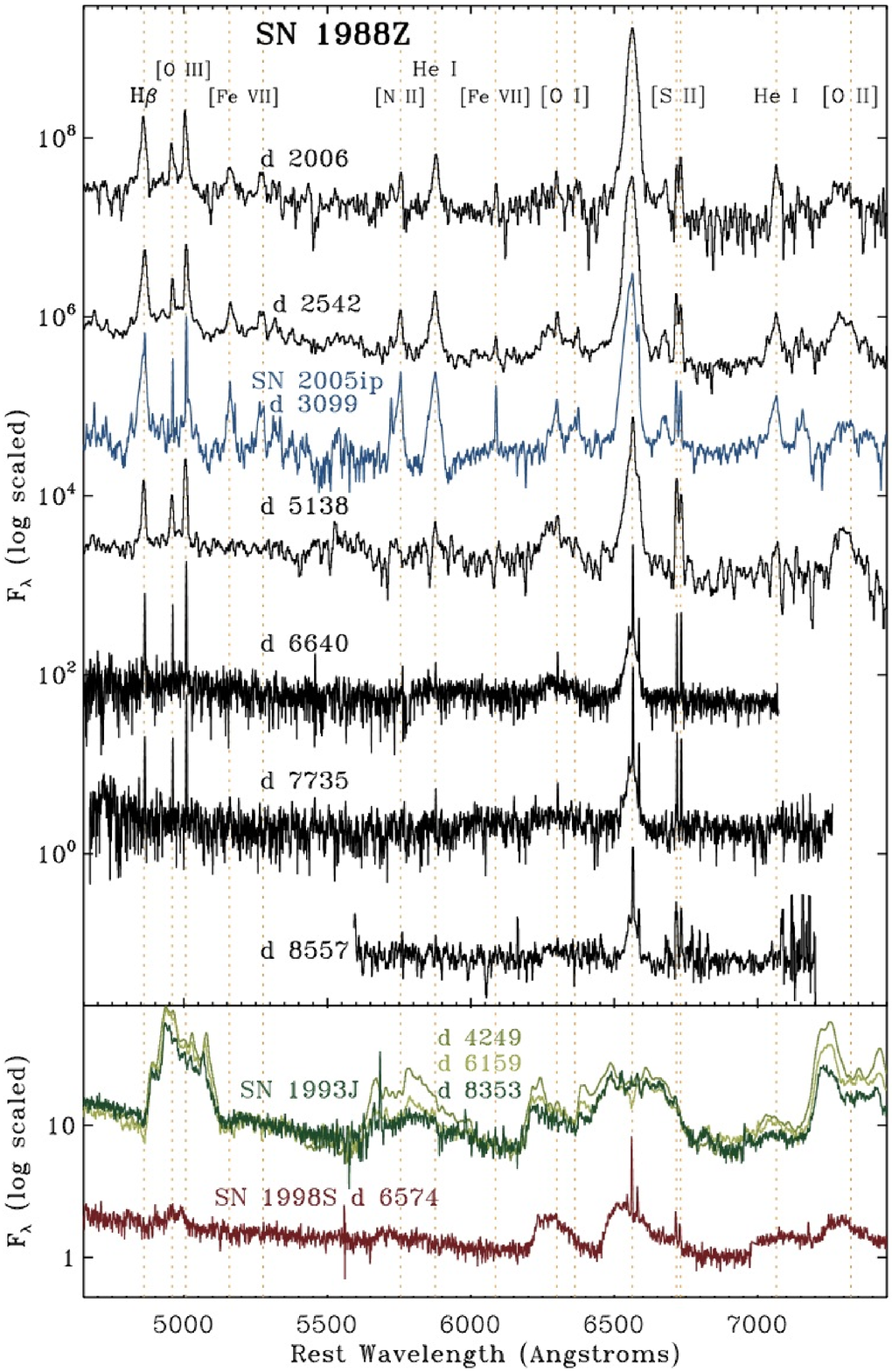}
\caption{The top panel shows late-time spectra of SN~1988Z (see
  Table~\ref{tab:88z}).  Vertical dashed orange lines identify
  specific emission lines.  We include the day 3099 spectrum of
  SN~2005ip (blue) for comparison. The bottom panel shows some of our
  additional late-time spectra of SN~1993J and SN~1998S.}
\label{fig:spectra88z}
\end{figure*}

\section{OBSERVATIONS}\label{sec:obs}

\subsection{Visual-Wavelength Spectroscopy}

We obtained late-time optical spectroscopy of SN~2005ip over several
epochs since our previous study \citep{smith09a}.  These observations
included the Bluechannel (BC) spectrograph on the 6.5\,m Multiple
Mirror Telescope (MMT), the Multi-Object Double Spectrograph (MODS;
\citealt{bo00}) on the Large Binocular Telescope (LBT),
%the Kast spectrograph \citep{1993ms} on the Lick 3\,m Shane reflector,
the Low-Resolution Imaging Spectrometer (LRIS; \citealt{oke95})
mounted on the 10\,m Keck~I telescope, and the Deep Imaging
Multi-Object Spectrograph (DEIMOS; \citealt{faber03}) on Keck~II.
Details of the spectral observations including measurements of the
total H$\alpha$ line flux are summarized in Table~\ref{tab:spec}, and
the spectra are shown in Figure~\ref{fig:spectra}. Standard reductions
were carried out using IRAF\footnote{IRAF, the Image Reduction and
  Analysis Facility, is distributed by the National Optical Astronomy
  Observatory, which is operated by the Association of Universities
  for Research in Astronomy (AURA), Inc., under cooperative agreement with
  the National Science Foundation (NSF).} including bias subtraction,
flat-fielding, and optimal spectral extraction (see, e.g.,
\citealt{kelson03}). Flux calibration was achieved using
spectrophotometric standards observed at similar airmass to that of
each science frame, and the resulting spectra were median combined into
a single one-dimensional spectrum.  For one observation noted in
Table~\ref{tab:spec}, the flux calibration was suspect owing to clouds
that came in during subsequent exposures.  For this date (day 3425) we
increase the uncertainty in Figure~\ref{fig:LHa} to $\pm50$\%.

% 88Z
{\it SN 1988Z:} In our study, we find it useful to compare the
evolution of SN~2005ip to that of a few other well-studied, geriatric
SNe with late-time interaction.  SN~1988Z is a prototypical,
long-lasting, and radio-luminous SN~IIn that has been studied in
detail by several authors
\citep{avf91,ss91,turatto93,vandyk+93,cd94,ft96,art99,williams+02,sp06}.
In particular, \citet{art99} have compiled multiwavelength data,
including H$\alpha$ fluxes, up to day $\sim$3000. We have obtained
additional unpublished late-time spectra of this aging SN~IIn using
the Keck Observatory, with both LRIS and DEIMOS. These observations,
including estimated H$\alpha$ line fluxes, are summarized in
Table~\ref{tab:88z} and almost triple the time baseline for this
object.  This series of new late-time spectra of SN~1988Z is shown in
Figure~\ref{fig:spectra88z}.  These spectra have been corrected for a
redshift of $z=0.022$, but no reddening correction has been applied as
in previous studies \citep{art99}.  When we plot the H$\alpha$ line
luminosity in Figure~\ref{fig:LHa}, we first subtract a constant flux
of $7 \times 10^{-16}$\,erg\,s$^{-1}$\,cm$^{-2}$, which is the mean value
of the narrow H$\alpha$ + [N~{\sc ii}] line fluxes in late-time
spectra (this is only significant for the last epochs).  We attribute
this constant narrow emission to contamination by coincident or nearby
H~{\sc ii} region emission, or perhaps unshocked distant CSM.  Since
we compare the H$\alpha$ luminosity of SN~2005ip to SN~1988Z and also
to other well-studied late interactors, we obtained late-time spectra
to extend the temporal coverage of two additional prototypical SNe.

{\it SN~1993J:} We procured several late-time spectra of SN~1993J,
most recently on 2016 Feb. 10 using LRIS on Keck I.  SN~1993J was a
Type Ib event rather than a Type IIn, but it was very nearby in M81,
was discovered within about 1 day of explosion, and has shown strong,
long-lived CSM interaction that has given rise to a high radio
luminosity and prominent optical/UV emission lines for more than a decade
after explosion \citep{fransson96,matheson00,bietenholz+02}.
Table~\ref{tab:93j} summarizes our late-time spectra reported here,
including a measurement of the total H$\alpha$ line flux in the
calibrated spectra.  Here the uncertainty is dominated by the fact
that the blue side of H$\alpha$ is blended with a strong oxygen line,
as well as uncertainty from possible slit losses.  Qualitatively,
the spectral appearance of SN~1993J changes very little at these late
epochs after day 4000, and even resembles the day 2454 spectrum
presented by \citet{matheson00}.  The most recent epoch is illustrated in
the bottom panel of Figure~\ref{fig:spectra88z} in dark green; we also
show two earlier epochs on days 4249 and 6159 in a lighter shade of
green for comparison.  Although some of the emission lines have
weakened with time, the line profiles and overall appearance of the
spectrum have not changed much.

{\it SN~1998S:} This has long been considered a prototypical SN~IIn,
which was discovered within about a day of explosion, but its early
CSM interaction was on the weak side compared to most SNe~IIn
\citep{leonard00,fassia01}.  SN~1998S did, however, exhibit strong
continued CSM interaction at late times \citep{pozzo04,ms12}.  We
obtained a late epoch of SN~1998S on 2016 Mar. 1 using DEIMOS on Keck
II, extending the spectral evolution by $\sim1500$\,d compared to
published data. The last previously published spectrum of SN~1998S
\citep{ms12} was obtained on day 5079, and our new spectrum
corresponds to day 6574. Qualitatively, the appearance of this
spectrum shows little change compared to the last one published by
\citet{ms12}.  For this spectrum, however, we measure a strong
H$\alpha$ line flux of $(2.15\pm1) \times
10^{-15}$\,erg\,s$^{-1}$\,cm$^{-2}$, which is several times higher
than our previous flux measurement \citep{ms12}.  The uncertainty is
likely to be dominated by absolute flux calibration and slit placement
as noted above for SN~2005ip and SN~1988Z, but this increase appears
to be significant.  The new spectrum of SN~1998S is shown in the
bottom panel of Figure~\ref{fig:spectra88z}.

\subsection{Radio}

We observed the position of SN~2005ip with the Karl G. Jansky Very
Large Array (VLA; Program 16A-101, PI Kilpatrick) starting on 2016
Feb. 4 (day 3743 after discovery) at 09:47:29 (UT dates are used
throughout this paper), at mean frequencies
of 6.2~GHz and $9.8$~GHz (with side-bands centred at 5.0, 7.4, 8.5,
and 11.0~GHz). We obtained 14\,min of on-source time for each
frequency, and used 3C286 and J0925+0019 for bandpass/flux and gain
calibration, respectively. Following standard procedures in the
Astronomical Image Processing System (AIPS; \citealt{gre03}) for data
calibration and analysis, we detect a bright source located at
$\alpha$ = \ra{09}{32}{06.41}, $\delta$ = $+$\dec{08}{26}{44.36}
(J2000; $\delta\alpha = 0.05''$, $\delta\delta=0.03''$), consistent
with the optical position. We measure flux densities and $1\sigma$
uncertainties for the upper and lower side-bands using AIPS/{\tt
  JMFIT}. The radio detections are listed in Table~\ref{tab:radio}.
The radio spectral index of SN 2005ip is $\alpha=0.55\pm0.11$, which
is consistent with optically thin emission observed from SNe~II
\citep{weiler+02}.  We postpone a more detailed analysis of the radio
data to a later paper that will examine its temporal variability.

%\begin{figure*}
%\vskip -0.0 true cm
%\centering
%\includegraphics[scale=0.56]{XRT_lc_countrate_2005ip.eps}
%\caption{This is not for the paper. For the paper we want a flux
%  calibrated light-curve, so that we can put together Swift and
%  Chandra on the same plot. In order to do that, I need a spectral
%  model and its evolution with time. Orange points: light-curve as
%  obtained by forcing the re-binning routine not to merge the last two
%  observations. Blue point: result of the merging of the latest two
%  observations. Statistically speaking, there is no difference. We
%  just need to decide what to show. Black dashed line: this is roughly
 % the level of the flux seen in your latest Chandra observation,
 % translated into XRT count-rate. Red points= points from Katsuda et
 % al. (you can see that at earlier timed they missed an observation,
 % nothing changes, but it is good to give the entire data set). Also:
 % I plotted here their count-rate as given in their table 1. From the
 % comparison from my extraction, I believe that those count-rates are
 % not corrected for PSF losses, while I plot corrected count-rates
 % (sometimes people do that, and give uncorrected count-rates and
 % corrected fluxes. I think this is the case here). We are basically
 % consistent (they also work in the 0.5-10 keV band, I used 0.3-10
 % keV).}
%\label{fig:XRTrate}
%\end{figure*}

\subsection{{\it Swift}-XRT}
\label{SubSec:XRTObs}

The X-Ray Telescope (XRT; \citealt{Burrows05}) onboard the
\emph{Swift} satellite \citep{Gehrels04} started observing SN~2005ip
on 2007 Feb. 14 (475 days since discovery). Late-time observations of
SN~2005ip have been carried out until 3630.5 days, as part of a {\it Swift}
fill-in program (PI Margutti).

\begin{figure}
\vskip -0.0 true cm
\centering
\includegraphics[width=3.3in]{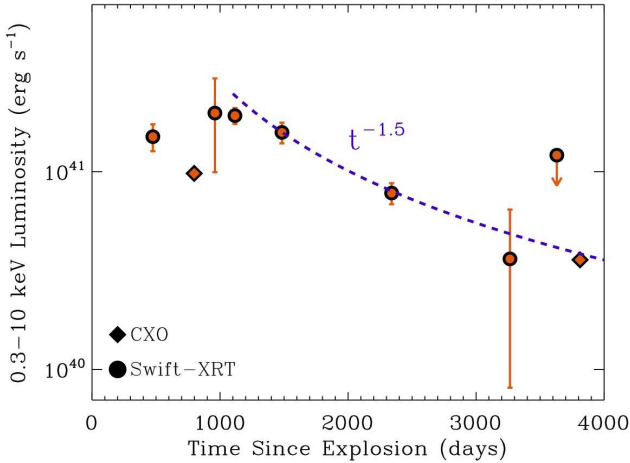}
\caption{X-ray luminosity evolution of SN~2005ip in the 0.3--10\,keV
  energy band. Circles refer to {\it Swift} data, and the diamonds are
  fluxes from {\it Chandra} data reported by \citep{katsuda14} as well
  as from our new {\it Chandra} observations (see text).}
\label{fig:05ipLx}
\end{figure}

XRT data have been analyzed using HEASOFT (v6.18) and corresponding
calibration files, following standard procedures (see
\citealt{Margutti13} for details). In particular, our rebinning
scheme requires a minimum of 20 photons in the source region and an
inferred source count rate at least $3\sigma$ above the background
level to yield a detection.  Compared to the previous compilation by
\cite{katsuda14}, our campaign extends the X-ray monitoring of
SN~2005ip from $\sim2400$ days to $\sim3600$ days since explosion,
and reveals a continuation of the fading X-ray emission with time.
The resulting luminosity light curve is shown in
Fig. \ref{fig:05ipLx}.

For the flux calibration of {\it Swift}-XRT data acquired at $t<2400$ days
we employ the spectral parameters of the one-component model derived
by \cite{katsuda14} (their Table 2) with a shock temperature evolution
$T\propto t^{-0.2}$. For {\it Swift}-XRT data acquired at $t>3000$ days we
adopt the spectral parameters constrained by our latest \emph{Chandra}
spectrum (see below).  Figure \ref{fig:05ipLx} shows the X-ray
evolution of SN~2005ip in the 0.3--10\,keV energy range as captured by
the {\it Swift}-XRT and {\it Chandra} in the first $\sim4000$ days since
explosion. The steep decline ($L_x\propto t^{-1.5}$) at late times
suggests that the shock is sampling a steeply decaying environment
density profile $\rho\propto r^{-s}$ with $s>2$ \citep{Fransson98}.

\begin{figure}
\vskip -0.0 true cm
\centering
\includegraphics[width=3.3in]{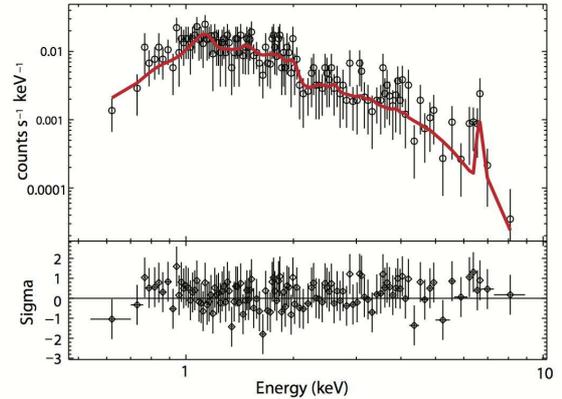}
\caption{\textit{Chandra} ACIS-S energy spectrum of SN~2005ip from
  2016 Apr. 3. The red curve is a thermal plasma model (see text), with
  model residuals shown in the lower panel.}
\label{fig:cxo_spec}
\end{figure}

\subsection{{\it Chandra} ACIS}
\label{SubSec:CXCObs}

SN~2005ip was observed with the \textit{Chandra X-ray Observatory}
(CXO) and Advanced CCD Imaging Spectrometer (ACIS) on 2016 Apr. 3 (3812
days) under a single-epoch DDT program (PI Mauerhan, ID 18802). The
ACIS-S array was used and the total on-source exposure time was 35.59
ks.

Photometry and energy spectra were extracted using the
\texttt{specextract} package within the HEASOFT \textit{Ciao} software
suite. The point source associated with SN~2005ip yielded 820 counts
within a circular aperture 2{\arcsec} in radius. A 5{\arcsec} radius
background annulus surrounding the source aperture yielded 121
counts. The energy spectrum of the source is shown in
Figure~\ref{fig:cxo_spec}. Note the apparent detection of Fe
K$\alpha$ emission near 6.8 keV. 

The source and background spectra were modeled simultaneously using
the \textit{Sherpa} package. An absorbed single-temperature thermal
plasma model (\textit{apec}) having solar abundances as defined by
\citet{Asplund09} was fit to the source for photon energies in the
range 0.5--8.0 keV, and a simple power law was used to fit to the
background. The He, C, N, O, and Fe elemental fractions, by number
relative to H, are $8.51\times10^{-2}$, $2.69\times10^{-4}$,
$6.76\times10^{-5}$, $4.90\times10^{-4}$, and $3.16\times10^{-5}$,
respectively.  A \textit{C}-statistic was used in the fitting.  We
fixed an equivalent interstellar neutral hydrogen column density of
$N_{\rm H} (\textrm{ISM})=3.7\times10^{20}$\,cm$^{-2}$, adopting this
value from \citet{katsuda14}, and allowed for an additional intrinsic
source of absorption for the SN. Our best fit yielded $N_{\rm H}
(\textrm{SN})=1.9_{-0.7}^{+0.8} \times 10^{21}$\,cm$^{-2}$ and a
plasma temperature of $kT=5.0_{-0.9}^{+1.1}$\,keV. The associated
uncertainties are 90\% confidence envelopes ($\sim1.6\sigma$). The
absorbed energy flux is $F_{\rm
  abs}=2.36_{-0.02}^{+0.06}\times10^{-13}$\,erg\,s$^{-1}$\,cm$^{-2}$
(0.5--8.0 keV) and the unabsorbed model flux is $F_{\rm
  unabs}=2.86_{-0.08}^{+0.12}\times10^{-13}$\,erg\,s$^{-1}$\,cm$^{-2}$. For
our adopted distance the flux implies an X-ray luminosity of
$L_X=3.2\times10^{40}$\,erg\,s$^{-1}$ for the SN at epoch 3812 days.
We also experimented with fitting the data using a ${\chi}^2$
statistic with a Gehrels variance function, and obtained a reduced
${\chi}^2$ value of 43.6 for 65 degrees of freedom. The resulting
physical parameters are consistent with those from the $C$ statistic,
quoted above. The model fit is also shown in
Figure~\ref{fig:cxo_spec}.

\begin{figure}
\includegraphics[width=3.0in]{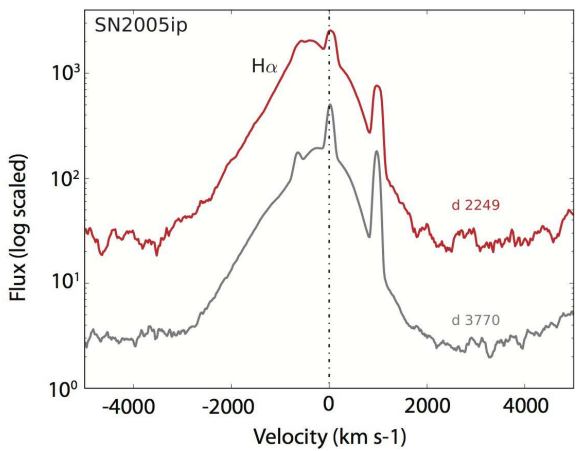}
\caption{A detail of the observed H$\alpha$ line profile, with
  velocity relative to the centroid of the narrow component.  This
  shows two of the higher-resolution spectra obtained with
  MMT/Bluechannel (day 2249) and Keck/DEIMOS (day 3770), before and
  after the dip in 2015, respectively.}
\label{fig:halpha}
\end{figure}

\begin{figure*}
\includegraphics[width=6.0in]{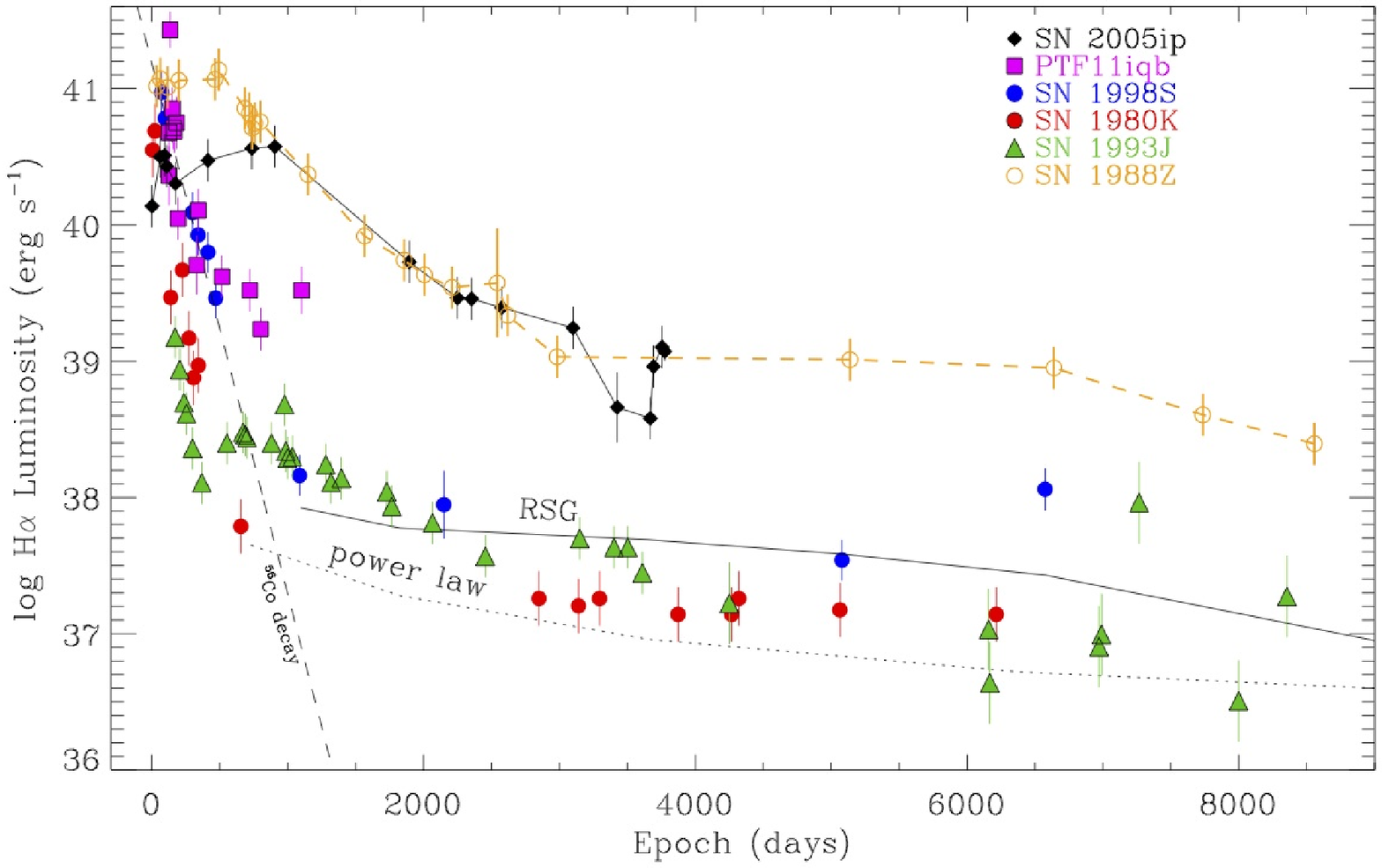}
\caption{Evolution of the total H$\alpha$ luminosity for SN~2005ip
  (black diamonds) compared to observations of other representative
  SNe with ongoing CSM interaction, as well as expected trends for a
  RSG wind profile \citep{cf94} and radioactive decay (dashed line).
  This figure is adapted from \citet{smith15}, which was adapted from
  \citet{ms12}.  H$\alpha$ luminosities for SN~1998S are from
  \citet{ms12}, SN~1980K from \citet{danny12}, SN~1993J from
  \citet{chandra09}, and PTF11iqb from \citet{smith15}.  For SN~1988Z,
  we use H$\alpha$ fluxes from \citet{turatto93} as described in our
  previous paper \citep{smith09a}, extended to almost day 3000 by
  \citet{art99}, plus our own measurements from unpublished spectra
  taken at later times of SN~1988Z (see Table~\ref{tab:88z}),
  SN~1993J, and SN~1998S.  The dashed black line shows the expected
  rate of fading for radioactive decay luminosity from $^{56}$Co, for
  comparison.}
\label{fig:LHa}
\end{figure*}

\section{RESULTS}\label{sec:res}

\subsection{H$\alpha$ Evolution of SN~2005ip}

Figure~\ref{fig:spectra} shows our newly obtained series of late-time
optical spectra of SN~2005ip, and Figure~\ref{fig:halpha} displays
the H$\alpha$ line profile in a few epochs with the best signal-to-noise 
ratio.  In addition to H$\alpha$, the spectrum shows consistent
emission from other lines commonly seen in late-time spectra of strongly
interacting SNe~IIn. Here we focus mainly on the H$\alpha$ line, since
it is the dominant line in the spectrum of SNe~IIn observed at late
times, and is a good proxy for the strength of interaction.

Our higher-resolution spectra (i.e., all but the 2014 May and 2015 Nov.
spectra) are able to cleanly resolve the intermediate-width emission
of the post-shock gas from the narrow pre-shock CSM or underlying
H~{\sc ii} region emission (Fig.~\ref{fig:halpha}).  In these spectra,
the intermediate-width component of H$\alpha$ (within about $\pm2000$
km s$^{-1}$) has a consistent profile shape, which is asymmetric and
shifted to the blue.  The peak of this emission is located roughly at
$-200$ to $-400$ km s$^{-1}$, shifting toward more positive velocities
with time.  This asymmetric blueshifted profile was seen at early
times in the broad and intermediate-width lines \citep{smith09a}, and
was attributed to new dust formation in the SN ejecta or in the
post-shock gas.  Blueshifted lines with this type of shape can arise
simply from occultation of the far side by the SN photosphere at early
times \citep{smith12,fransson14,dessart15}, but the fact that the
blueshifted asymmetry persists for a decade in SN~2005ip, long after
the continuum opacity of the photosphere has vanished, means that the
blocking of the red side of the lines must be caused by dust mixed within
the SN (or to intrinsically asymmetric CSM).  Much of the dust is
likely to be mixed in the post-shock region where the
intermediate-width H$\alpha$ originates, since even early-time spectra
showed a blueshift in lines formed within the post-shock shell, as
well as in the broad SN ejecta lines \citep{smith09a}.  The narrowest
components from the pre-shock CSM are unresolved or only marginally
resolved in our spectra, indicating expansion speeds of 80 km s$^{-1}$
or less (Fig.~\ref{fig:halpha}).

The temporal evolution of the H$\alpha$ line luminosity is shown in
Figure~\ref{fig:LHa}. This plot uses H$\alpha$ line fluxes in the
first 1000 d from our earlier paper \citep{smith09a}, and supplements
these with our new spectra of SN~2005ip reported here.  The line
fluxes were converted to H$\alpha$ luminosity with the same
assumptions as in our earlier study.  The H$\alpha$ luminosity remains
roughly constant for the first 1000 d, as we noted earlier, and then
declines slowly over the next several years as found previously by
\citet{stritz12}.  Our new spectra show that this slow decline
continued to around day $\sim3600$ (mid 2015), when the H$\alpha$
luminosity was in decline, and then suddenly reversed to be on a quick
rise that has continued to the time of writing. Additional monitoring
will be needed to determine how high it will rise, if it levels off,
or fades again.  Possible causes of this are discussed further in
Section 4.

Figure~\ref{fig:LHa} also compares the H$\alpha$ luminosity of
SN~2005ip to that of several other SNe with strong late-time
interaction.  We see here that SN~2005ip really is quite unusual, with
a sustained H$\alpha$ luminosity that was essentially the same as the
current record holder for such interaction: SN~1988Z (see below).  At
days 1000--2000, the H$\alpha$ luminosity was 2 orders of magnitude
stronger than in some other well-studied objects like SN~1998S
\citep{ms12}, SN~1980K \citep{danny12}, and SN~1993J
\citep{chandra09}, and an order of magnitude stronger than in the
SN~IIn PTF11iqb \citep{smith15}. The late-time CSM interaction in
these other objects can be explained by a very strong RSG wind with
$10^{-4}$ M$_{\odot}$ yr$^{-1}$ \Citep{ms12}, but SN~2005ip and
SN~1988Z are clearly more extreme situations.  Thus, the H$\alpha$
luminosity implies that around 4000 days after explosion, the shock is
interacting with a wind that had a mass-loss rate of at least
$10^{-3}$ M$_{\odot}$ yr$^{-1}$.  With a CSM speed\footnote{This is a
  guess, but is lower than the 80 km s$^{-1}$ resolution limit of our
  data, and similar to CSM speeds for SN~1998S seen in echelle spectra
  \citep{shivvers15}.} around 40 km s$^{-1}$, this implies that the
high mass-loss rate was occurring within a few thousand years before
core collapse. This, in turn, requires several M$_{\odot}$ being lost
from the star in the last few millennia of its life. Our estimate is
conservative compared with some others, which favour higher rates of
$\ga 0.01$ M$_{\odot}$ yr$^{-1}$ and total CSM masses greater than 10
M$_{\odot}$ \citep{katsuda14,stritz12}.

In our earlier paper \citep{smith09a}, we drew comparisons between
SN~2005ip and SN~1988Z, including a direct comparison of the H$\alpha$
luminosity.  During the first several hundred days, SN~2005ip was
significantly less luminous in both H$\alpha$ and continuum emission.
However, during this time SN~2005ip remained roughly constant in
H$\alpha$ luminosity, while SN~1988Z steadily declined.
Figure~\ref{fig:LHa} extends this same comparison, and shows that by
day 1000 the two objects had about the same luminosity, and for the
rest of its first decade, SN~2005ip was an almost exact twin of
SN~1988Z in terms of its CSM interaction strength traced by H$\alpha$.
As in our previous paper, Figure~\ref{fig:LHa} incorporates H$\alpha$
luminosities for SN~1988Z published by \citet{turatto93}, extended to
almost day 3000 by \citet{art99}.  We also supplement these published
H$\alpha$ luminosities with measurements from our own spectra of
SN~1988Z (see Table~\ref{tab:88z}).

With its recent resurgence in H$\alpha$ luminosity, SN~2005ip equals
SN~1988Z again at a comparable epoch, making it tied as the
record holder for the highest late-time H$\alpha$ luminosity at +10\,yr. 
Integrating through time in Figure~\ref{fig:LHa}, the total energy
radiated by SN~2005ip in the H$\alpha$ line alone is
$4 \times 10^{48}$\,erg.  For comparison, the total integrated 
0.3--10\,keV X-ray energy is about a factor of 10 larger at
$\sim4\times10^{49}$\,erg, and the integrated optical luminosity
yields a similar total radiated energy of $\sim4\times10^{49}$\,erg.  
This is certainly an underestimate of the total radiated
energy, since we appled no correction for flux at other wavelengths.
The asymmetric blueshifted profiles of H$\alpha$, in particular,
suggest some dust extinction in SN~2005ip (with a lack of such
evidence for dust extinction in SN~1988Z); correcting for this could
raise the H$\alpha$ and continuum luminosities further, and thus make
SN~2005ip significantly more luminous than SN~1988Z.  The radiated
energy so far is therefore likely to be at least $10^{50}$\,erg for
SN~2005ip.  The kinetic energy imparted to the swept-up CSM shell
(about 10\,M$_{\odot}$ accelerated to $\sim2000$\,km\,s$^{-1}$) is
roughly $4 \times 10^{50}$\,erg.  Thus, CSM interaction may have
already tapped a substantial fraction of the explosion energy in
SN~2005ip; remaining kinetic energy of the freely expanding ejecta
that are still inside the reverse shock may power SN~2005ip for the
next decade.  It will be interesting to see where it goes from here.
A significant detail is that after it rebrightened, the spectrum
showed stronger emission from the intermediate-width components of
[O~{\sc i}] $\lambda\lambda$6300, 6364, which had only shown narrow
pre-shock CSM emission before that time. This may hint that we are
starting to see an increased contribution from the reverse-shock
luminosity (see Section 3.3 below).

\subsection{Very Late-time Evolution of  SN~1988Z}

Despite its redshift of 0.022, SN~1988Z has remained detectable for
nearly three decades.  It was a very luminous radio SN that has been
studied extensively as it faded slowly after discovery, showing
remarkable longevity in the radio, X-rays, and optical emission lines
like H$\alpha$
\citep{avf91,ss91,turatto93,vandyk+93,cd94,ft96,art99,williams+02,sp06}.
It has been compared to SN~1986J, and most studies of its first decade
and a half favour the interpretation that it is powered by an
energetic explosion (as much as $\sim10^{52}$\,erg; \citealt{art99})
from a massive progenitor (an initial mass of something like 
20--30\,M$_{\odot}$ or more), that was interacting with roughly 
10\,M$_{\odot}$ produced by a star that had a wind with a mass-loss rate 
of $\sim10^{-3}$\,M$_{\odot}$\,yr$^{-1}$ for $\sim10^4$\,yr before core
collapse.  The rate of decline in H$\alpha$ and radio suggested that
the mass-loss rate ramped up in the final millennium before core
collapse \citep{vandyk+93,art99,williams+02}.  Even the somewhat lower
mass-loss rate traced by later CSM interaction was extreme, however,
allowing SN~1988Z to remain as one of the most radio luminous SNe even
after a decade, matched only by SN~2005ip.

Here we present a series of optical spectra, listed in
Table~\ref{tab:88z} and shown in Figure~\ref{fig:spectra88z}, which
extend this late-time evolution from the mid-1990s to the present
epoch. We find that SN~1988Z's strong CSM interaction has continued to
fade very slowly, indicating a remarkably extended and dense wind into
which the blast wave continues to crash.  This traces mass loss over
$\sim10^{4}$\,yr preceeding core collapse.  Based on the luminosity
of H$\alpha$ and the character of the optical spectra, we group
SN~1988Z's late evolution into a few different epochs:

(1) {\it The first 500--3000 days}.  This is the phase that has already
been studied extensively as noted above.  During this time, SN~1988Z
showed a steady and very slow decline in radio and H$\alpha$
luminosity, although remaining far more luminous than most SNe having
signs of CSM interaction.  \cite{williams+02} noted a possible break
in the decline rate of the H$\alpha$ luminosity around day 1000,
although this break is not clear in Figure~\ref{fig:LHa} where the
decline looks rather continuous within the uncertainties.  Of interest
in the present paper is that throughout the time period of roughly
1000--3000 days, SN~2005ip was a nearly identical twin of SN~1988Z in
terms of its decline rate, its luminosity in the radio (admittedly only a
single epoch so far for SN~2005ip), H$\alpha$, and X-rays, and in
the appearance of its spectrum.  In our previous paper
\citep{smith09a} we noted similarities in the spectra of SN~2005ip and
SN~1988Z at early times, although SN~2005ip exhibited stronger coronal
lines.  From day 1000 to 3000, however, their spectra are almost
indistinguishable.  Figure~\ref{fig:spectra88z} includes a recent
spectrum of SN~2005ip on day 3098 for comparison, which appears very
similar to the days 2006 and 2542 spectra of SN~1988Z in the same
plot.  This phase is characterized by very strong intermediate-width
H$\alpha$, a steep H$\alpha$/H$\beta$ decrement of $>10$, several
intermediate-width and narrow coronal lines such as [Fe~{\sc vii}],
and intermediate-width lines that indicate very high electron
densities in the post-shock gas, such as [N~{\sc ii}] $\lambda$5755
(which is, however, blended with an [Fe~{\sc vii}] line).

(2) {\it A ``plateau'' over days 3000--7000}.  During this time
period, the H$\alpha$ luminosity appears to level-off for a decade.
While H$\alpha$ remains constant, the rest of the spectrum shows some
interesting changes.  The coronal lines and other indicators of high
density fade away, leaving mostly narrow H~{\sc ii} region lines and
broad, weak oxygen lines (possibly from the reverse shock; see below)
that are also seen in young SN remnants.  The intermediate-width
H$\beta$ line also fades, yielding an H$\alpha$/H$\beta$ ratio that is
higher than before ($>30$).

(3) {\it After day 7000.} Eventually, other lines fade away too,
leaving only intermediate-width H$\alpha$, which also resumed a
decline in flux at these very late times.  Even the
broad/intermediate-width oxygen lines mostly disappear.  Note that in
Figure~\ref{fig:spectra88z}, we have subtracted a constant value for
the contribution of narrow H$\alpha$+[N~{\sc ii}] emission from
underlying H~{\sc ii} regions in the spectrum, which typically
contributes about $7 \times 10^{-16}$\,erg\,s$^{-1}$\,cm$^{-2}$ in our
Keck slit aperture after subtraction of adjacent background.  This
suggests that SN~1988Z resides in or near a complex of H~{\sc ii}
regions, providing another possible indication of a high initial mass
for the progenitor.  At SN~1988Z's distance of $\sim 100$\,Mpc,
however, $1''$ corresponds to almost 500\,pc, which could include a
large complex of H~{\sc ii} regions like the Tarantula nebula, and is
not a very reliable or precise indicator of initial mass.  It is,
however, a good indicator of a relatively strong ambient UV radiation
field.

This last point is interesting in the context of something we discuss
later.  Namely, \citet{mackey14} have suggested that external
ionization of an otherwise normal RSG wind can produce a stalled dense
shell, which might offer an alternative to very strong or eruptive
mass loss as an origin for SNe~IIn.\footnote{Note that this scenario
  cannot explain the CSM interaction observed at early times in normal
  SNe~IIn, because their narrow emission line components have resolved
  widths and P-Cygni absorption indicating that the CSM near the star
  is {\it outflowing} --- i.e., the CSM shells are not stalled and are
  not ambient material (see \citealt{smith14}). Stalled shells of this
  sort, however, could potentially influence the shock interaction
  observed at very late times.}  Despite its probable location in an
H~{\sc ii} region, this idea does not seem to apply well to SN~1988Z,
which from its earliest phases shows an uninterrupted decline in CSM
interaction strength, consistent with a strong shock moving through a
freely expanding wind.  The total mass of 15--20\,M$_{\odot}$ or more
of CSM that has been swept up so far by SN~1988Z rules out a normal
RSG wind that has been confined to a thin shell, because such shells
only contain a fraction (about a third) of the total RSG mass loss,
and are therefore expected to have much lower total mass
\citep{mackey14}.  With such a strong wind, the expected location of a
stalled shell would probably be different from the models presented by
\citet{mackey14}, however.  We return to this issue later, because
despite its similarity to SN~1988Z, SN~2005ip does show a rapid
interruption of the smooth decline in its CSM interaction strength,
which might be indicative of external influence of this sort.  This
may hint that two otherwise very similar events may have their CSM
modified in different ways.  Then again, the relatively coarse
sampling of SN~1988Z in its past decade does not rule out the
possibility of some brief drops or spikes in luminosity indicative of
density fluctuations in its CSM.  Indeed, one may speculate that the
late plateau phase in SN~1988Z's H$\alpha$ luminosity might represent
the shock traversing a relatively constant density, ionized portion of
the wind outside such a neutral dense shell.  For this reason as well,
it will be interesting to see how SN~2005ip behaves in the coming
decade as compared to SN~1988Z.

\subsection{Line Profiles of Elderly Interactors}

All the SNe discussed in this paper show signs of long-lived, strong
CSM interaction, but there are interesting differences.  Examining
Figures~\ref{fig:spectra} and \ref{fig:spectra88z}, it is clear that
both SN~2005ip and SN~1988Z have late-time spectra dominated by their
intermediate-width components of H$\alpha$ and other lines, with line
widths of about 2000\,km\,s$^{-1}$.  Both SN~1993J and SN~1998S have
significantly broader line profiles of 5000--10,000\,km\,s$^{-1}$.  This
difference is probably closely linked to their total H$\alpha$
luminosity.  Both SN~2005ip and SN~1988Z are $\sim2$ orders of
magnitude more luminous than the other two at a comparable late epoch.
The combination of higher H$\alpha$ luminosity (as well as higher
radio and X-ray luminosities), high-density tracers, and narrower
lines likely stems from the same cause: {\it denser CSM and a
  radiative forward shock}.  If a blast wave is expanding into
extremely dense and extended CSM, the forward shock can remain
radiative, giving higher CSM interaction luminosity and causing the
intermediate-width components to continue to dominate the spectrum as
the blast wave continues to expand into the CSM and to decelerate
slowly.  (Strong global asymmetry or a high degree of clumping can
provide very high CSM densities without necessarily having a huge
total CSM mass, so an intermediate-width component might also be
present at lower luminosity in some cases.)

At lower average CSM densities and luminosities, as in SN~1993J and
SN~1998S, radiation from the reverse shock dominates the late-time
emission, and so the spectrum exhibits much broader lines from
oxygen-rich SN ejecta crossing the reverse shock.  In our late spectra
of SN~1993J, the flat-topped and double-peaked profiles described by
\citet{matheson00} still persist. These may be the result of
interaction with a disk-like CSM, which pinches the waist of the CSM
interaction region and makes the reverse shock brighter at the
equator.  One may envision a reverse-shock geometry similar to that of
SN~1987A \citep{france15}.

Interestingly, we may see a transitional phase in SN~1988Z, where weak
and broad oxygen lines remain after the intermediate-width coronal
lines have faded (days 5138 and 6640 in Figure~\ref{fig:spectra88z}).
In general, though, the lack of strong oxygen lines in most SNe~IIn at
a few hundred days after explosion is most likely a consequence of the
fact that the forward shock still dominates the emission spectrum.
Since this is tracing shock-heated CSM material, it reflects the
composition of the CSM and not the composition of the inner ejecta.
Thus, a lack of strong oxygen features should not be taken as any
indication of the abundances in the SN ejecta as long as
intermediate-width lines are still seen.

Another relevant detail concerns the line profiles.  While SN~2005ip
and SN~1988Z are otherwise very similar, they differ in that SN~2005ip
shows asymmetric blueshifted line profiles, especially in its
intermediate-width H$\alpha$ line at late times
(Figure~\ref{fig:spectra}).  The intermediate-width H$\alpha$
components of SN~1988Z remain much more symmetric at all epochs
(Figure~\ref{fig:spectra88z}).  This could signify a difference in
dust-formation efficiency in the two SNe, or perhaps viewing-angle
effects if the CSM interaction is not spherically symmetric (for
example, SN~1988Z could be a ring or disk-like interaction seen
pole-on, in which case dust cannot block emission from receding
material).  Interestingly, the late-time line profiles of SN~1998S,
and to a somewhat lesser extent SN~1993J, do show blueshifted
asymmetry that may be indicative of dust formation in these SNe.  This
was already discussed in detail for the case of SN~1998S by
\citet{ms12}.  There have been no published studies of early-time
spectropolarimetry of SN~1988Z or SN~2005ip, however.

\begin{figure}
\includegraphics[width=3.5in]{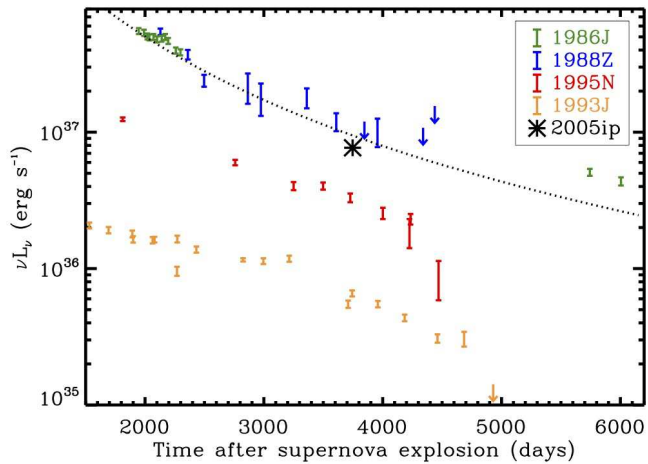}
\caption{Late-time (1500--6000 days) 5~GHz luminosity
  observed toward SNe 1986J, 1988Z, 1995N, and 2005ip \citep[$d = 10$,
  100, 24, 33 Mpc,
  respectively;][]{williams+02,bietenholz+02,chandra09,smith09a}.  We
  have assumed that the radio emission from each event is emitted
  isotropically.  The dotted curve shows the best-fit model to
  late-time ($>1500$~day) emission from SN~1988Z.}
\label{fig:radio}
\end{figure}

\subsection{Radio Emission from SN~2005ip}

Very few radio SNe have been observed beyond 10~yr.
Previous examples such as SNe 1986J, 1988Z, and 1995N \citep[see,
e.g.,][]{vandyk+93,williams+02,bietenholz+02,chandra09} were noted for
having unusually luminous radio emission that peaked around 900 to
1100~days after the inferred explosion date.  In
\autoref{fig:radio}, we compare the 5~GHz radio luminosity
observed toward SN 2005ip to these radio SNe observed at late times
(1500--6000~days).  SN~2005ip appears very similar in its
radio luminosity to SN~1988Z at a comparable epoch, which is perhaps
not surprising in light of the H$\alpha$ results above.  The best fit
to the observed radio light curve of SN~1988Z is based on a
five-parameter model derived from \citet{weiler+86}:

\begin{eqnarray}
	S_{\nu} &=& K_{1} \left(\frac{\nu}{5~\text{GHz}}\right)^{\alpha} \left(\frac{t-t_{0}}{1~\text{day}}\right)^{\beta} \left(\frac{1 - e^{\tau}}{\tau}\right), \\
	\tau &=& K_{2} \left(\frac{\nu}{5~\text{GHz}}\right)^{-2.1} \left(\frac{t-t_{0}}{1~\text{day}}\right)^{\delta},
\end{eqnarray}

\noindent with $\alpha, \beta, \delta, K_{1}$, and $K_{2}$ as free
parameters.  The explosion date $t_{0}$ can be fixed from early-time
optical photometry.  In \citet{williams+02}, the fit parameters for
the late-time ($>1500$~day) light curve of SN 1988Z were
$\alpha = -0.72$, $\beta= -2.73$, $\delta = -2.87$, $K_{1} = 9.1 \times
10^{8}$, and $K_{2} = 3.19 \times 10^{8}$.  These parameters imply that
the SN progenitor underwent mass loss of at least $1.2 \times
10^{-4}$\,M$_{\odot}$\,yr$^{-1}$ for the last
10,000~yr before exploding.  SN~2005ip is about 20\% less
luminous than SN 1988Z at 5~GHz and $t \approx
3750$ days, but it is a factor of a few more luminous in X-rays
at the same epoch.  Analysis of X-rays suggests significantly larger
mass-loss rates of $10^{-3}$ to $10^{-2}$\,M$_{\odot}$\,yr$^{-1}$
\citep{katsuda14}.

It is difficult to infer the physical parameters of the CSM
surrounding SN 2005ip based on a single epoch of radio observations.
However, given the similarity between SNe~2005ip and 1988Z at both
optical and radio wavelengths, we can hypothesize that
SN~2005ip has an extremely dense and clumpy CSM profile out to large
radii.  We plan to continue to monitor the radio and X-ray evolution
of SN~2005ip in light of its recent H$\alpha$ brightening, and the
results will be presented in a future paper.

%\begin{figure}
%\includegraphics[width=3.5in]{art99xray.eps}
%\caption{Late-time X-ray luminosity of SN~2005ip (from {\it Swift} and
%  {\it Chandra} observations reported here, compared to the late-time
%  X-ray luminosity of SN~1988Z at a similar epoch \citep{art99}.  This
%  is obviously a place holder (Xrays from Aretxaga 1999).  Need a new
%  fig similar to radio.}
%\label{fig:LX}
%\end{figure}

\begin{figure}
\vskip -0.0 true cm
\centering
\includegraphics[width=3.5in]{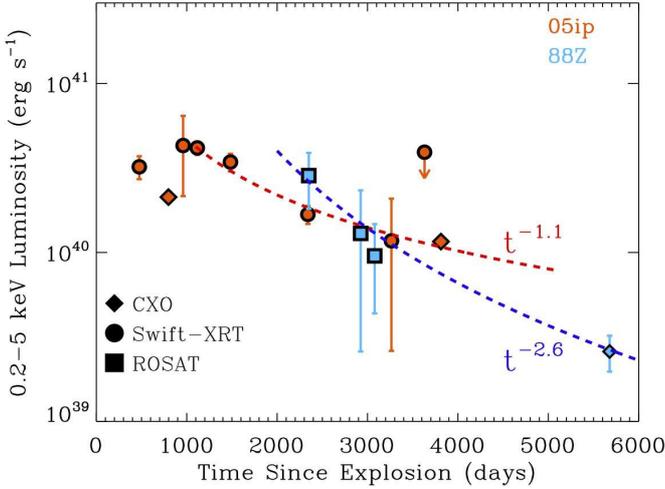}
\caption{X-ray evolution of SN~2005ip (this paper) and SN~1988Z
  (\citealt{sp06}). The X-ray luminosity plotted for SN~1988Z is a
  lower limit to the true luminosity of the transient, as no
  correction for any intrinsic neutral hydrogen column density was
  attempted by \citet{sp06}. For SN~1988Z \citet{sp06} assumed a 5\,keV 
  plasma in collisional equilibrium.  For comparison with this
  X-ray light curve of SN~1988Z, here we show the X-ray fluxes for
  SN~2005ip reduced in the same way (not corrected for intrinsic
  absorption, integrated over the same energy range of 0.5--2\,keV).
  This is different from the data shown in Figure~\ref{fig:05ipLx},
  but allows for a more direct comparison with SN~1988Z.  The two SNe
  overlap in X-ray luminosity in this energy band, but have somewhat
  different deecay rates.}
\label{fig:05ip88Z}
\end{figure}

\subsection{Late-Time X-ray Emission}

As noted in Section 2.3, the rate of fading of the X-ray luminosity of
SN~2005ip implies a CSM density that falls off more steeply than for a
steady wind with $\rho \propto r^{-s}$ and $s=2$, which should yield a
$t^{-1}$ profile in the case of free-free emission from a wind.  This
would imply that the star's mass-loss rate was ramping up in the last
few thousand years of its life as it approached core collapse.  This
is qualitatively consistent with implications from the H$\alpha$
evolution noted above.  The wind density being steeper than a steady
wind is also consistent with the earlier analysis of X-rays by
\citet{katsuda14}.  Looking at X-rays alone, the rate of fading
appears to be consistent with a continuation of the trend inferred by
\citet{katsuda14}.  This is interesting, since the H$\alpha$
luminosity shows a more sudden drop followed by a resurgence in recent
data.  It is possible that the coarse time sampling of the X-ray data
has missed this peculiar variability, so future additional X-ray
observations are required to resolve this.

Although the X-rays from SN~2005ip imply a CSM that falls off more
steeply than a steady wind, it is interesting to note that the X-ray
luminosity of SN~1988Z falls off even faster with time at a similar
epoch after explosion than SN~2005ip (e.g., Figure~\ref{fig:05ip88Z}
and \citealt{sp06}).  Note that a direct comparison between X-rays
from SN~2005ip and SN~1988Z is a bit tricky, because \citet{sp06} made
somewhat different assumptions in their analysis.  Therefore, for the
purpose of comparing the two objects in Fig~\ref{fig:05ip88Z}, we have
recalibrated our SN~2005ip X-ray data with the same assumptions
(integrated over 0.5--2\,keV, uncorrected for intrinsic absorption).
Here we find that the X-ray luminosities are about the same at
$\sim3000$\,d, although the somewhat faster decline of SN~2005ip is
still apparent.  Yet, despite this faster decline in X-rays around a
decade after explosion, the H$\alpha$ luminosity of SN~1988Z then hit
a plateau and did not decline for another decade after this epoch.
This constant H$\alpha$ luminosity seems inconsistent with the steep
drop in wind density implied by the X-rays.  Overall, it seems that a
clear lesson from comparing various diagnostics from different
late-time interactors is that any single diagnostic by itself provides
an incomplete picture, so readers should be aware that some of the
corresponding mass-loss rate estimates are likely to be lower limits.
Note that the value of $\dot{M}$ that we derive from radio emission is
an order of magnitude lower than from X-ray and H$\alpha$ observations
of SN~2005ip at the same epoch.

Consider the comparison of SN~1988Z and SN~2005ip a decade after
explosion.  The two have roughly equal X-ray and H$\alpha$
luminosities, but SN~1988Z is more luminous in radio emission.  A
caveat is that the quoted X-ray luminosities are not bolometric X-ray
luminosities, although we have tried to recalibrate our SN~2005ip
X-ray data in the same way as had been done for SN~1988Z \citep{sp06}.
One might imagine that these inconsistencies could be attributed to
some combination of clumping or aspherical geometry, which can
influence the emissivity of H$\alpha$ or the escape of X-rays and
radio differently for a given CSM mass.  Dense regions that emit
strong H$\alpha$ emission might be self absorbed in the radio, for
example, and a range of densities may exist simultaneously if the
emitting region is clumpy.  \citet{katsuda16} discussed how asymmetry
might influence X-ray emission in SNe~IIn.  We noted earlier that
there are some indications of asymmetry and different viewing angles
between the two SNe (SN~2005ip shows asymmetric blueshifted line
profiles in H$\alpha$ that imply dust formation blocks the far side,
whereas SN~1988Z does not; this could perhaps be ascribed to a pole-on 
view of SN~1988Z and a nearly edge-on view of SN~2005ip).  While
such aspects are difficult to constrain with high confidence, the
large binary fraction among massive stars
\citep{moe16,sana12,chini12,kk12,kiminki12,chip14} and the high
incidence of aspherical geometry in resolved CSM around nearby massive
stars suggests that effects such as asymmetry and clumping may be the
norm rather than an exception.

\section{DISCUSSION}\label{sec:disc}

In this paper we have emphasized the remarkable similarity between the
late-time evolution of SN~2005ip and that of SN~1988Z, which is the
prototypical long-lived SN~IIn.  So far there are three main
differences in their evolution.  

First, the CSM interaction in SN~1988Z was stronger at early times and
declined for the first 1000 days, whereas SN~2005ip had a more delayed
onset of its strongest CSM interaction.  SN~2005ip's interaction
remained roughly constant until day 1000, after which it tracked that of
SN~1988Z.  This relatively delayed onset in SN~2005ip was accompanied
by stronger narrow coronal emission lines at early times, probably
signifying a more highly clumped inner-wind region (compared to
SN~1988Z), allowing the densest pre-shock regions to be photoionized
by X-rays from a very fast blast wave \citep{smith09a}.  

The second major difference is that SN~2005ip has just recently shown
a drop and then a quick resurgence in its H$\alpha$ luminosity over
the course of a few months, indicative of strong density fluctuations
in its distant CSM.  The recent sudden increase in H$\alpha$ would
reflect a density increase by a factor of 2--5.  SN~1988Z has shown no
such behaviour, although brief blips cannot be ruled out by the coarse
time sampling of our late-time SN~1988Z spectra.  For a blast wave
expanding at 2000--5000\,km\,s$^{-1}$, fluctuations seen $\sim3500$ days
after SN~2005ip's explosion would correspond to radii of (0.6--1.5)
$\times 10^{17}$\,cm, or roughly 0.02--0.05\,pc from the star.
Interestingly, this is consistent with the radius of a warm dust shell
responsible for the near/mid-IR echo seen at early times
\citep{ori1,ori2}, which was inferred to reside at $\sim0.03$\,pc from
the star \citep{fox11,fox13}.  This dust was ascribed to a shell of
enhanced density at that location.  It therefere seems likely that the
sudden increase in the H$\alpha$ luminosity that we observe around day
3500 could in fact be caused by the forward shock reaching and
overtaking this same dusty CSM shell.  If so, continued study of this
interaction may provide a novel way to investigate the destruction or
survival of CSM dust that is hit by a SN blast wave.

Third, despite their similar H$\alpha$ luminosity evolution, SN~2005ip
declines more slowly in X-rays than SN~1988Z, but it is also somewhat
less luminous at radio wavelengths than SN~1988Z at a comparable epoch
(see Sections 3.4 and 3.5 for details and caveats). This suggests that
any one diagnostic taken alone gives an incomplete picture of the
progenitor mass loss.

The recent H$\alpha$ variability suggests that SN~2005ip has strong
deviations from a monotonically decreasing wind density at large
radii, but what caused this?  A simple interpretation would be that
changes in CSM density reflect previous changes in wind mass-loss
rate.  At a radius of 0.03\,pc, and with a constant wind speed of
$\sim40$\,km\,s$^{-1}$ (somewhat faster wind speeds are appropriate for
more luminous RSGs; see \citealt{smith14}), this would point to
substantial changes in mass loss about 1000\,yr preceding core
collapse. As noted in Section 1, sudden bursts of mass loss on these
timescales before core collapse do not fall within the realm of
currently suggested ideas that involve Ne and O burning, which are too
brief for objects like SN~2005ip and SN~1988Z \citep{qs12,sq14,sa14}.
Instead, it appears as if the mass-loss rate was ramping up in the
last few thousand years before core collapse in SN~2005ip and
SN~1988Z.  Some evidence for such behaviour is observed among nearby
populations of RSGs \citep{bd16}, and it may be expected theoretically
\citep{heger97,yc10,smith14}.

Although RSG winds may ramp up near the end of a star's life, strong
fluctuations in CSM density may point to episodic mass loss akin to
eruptions of luminous blue variables (LBVs) or unsteady binary
mass-transfer episodes.  Indeed, LBV-like eruptive mass loss has been
suggested as the possible origin of the CSM shells around SN~2005ip
\citep{fox13,katsuda14}.  \citet{sa14} have discussed how binary
interaction triggered by pre-SN evolution might instigate LBV-like
mass loss through a rapid onset of common-envelope evolution.
Recalling that LBVs are a phenomenological class and binary
interaction is a physical mechanism, these might be two names for the
same phenomenon.  Some current ideas for LBVs do favour binary
interaction as a necessary ingredient in their evolution \citep{st15}.

A drop in pre-shock density followed immediately by an increase in
density, as indicated by the H$\alpha$ fluctuation of SN~2005ip, might
also result from a temporary increase in wind speed that swept
material into a thinner shell.  Thus, variations in wind speed, as
opposed to just wind mass-loss rate, might also give rise to such
structures.  Wind speeds could vary significantly if the progenitor
experienced a blue loop on the Hertzsprung-Russell diagram, 
for example. Perhaps the
common assumption of ballistic speeds is naive, but nevertheless, this
material must be far from the star and quite old compared to the CSM
overtaken by normal SNe~IIn during their early bright phases.

Another puzzling aspect of SN~2005ip's late-time behaviour concerns its
X-ray emission.  In 2013--2014, analysis of X-rays from SN~2005ip
\citep{katsuda14} indicated a drop in both the X-ray flux and the
absorbing neutral H column density ($N_{\rm H}$) along our line of sight
(reaching the expected Galactic line-of-sight value by the last
epoch).  This prompted \citet{katsuda14} to suggest that SN~2005ip's
blast wave had reached the outer boundary of the dense shell.
However, the similarity to SN~1988Z (which continued unabated for
another decade after this epoch), the recent resurgence in H$\alpha$
emission from SN~2005ip, and strong radio and X-ray emission, indicate 
that its CSM interaction is not yet going away.  In that case, how should
we reconcile the observed drop in $N_{\rm H}$?  \citet{katsuda14}
considered the possibility that the outer RSG wind is ionized by the
luminosity from CSM interaction, but found the ionization to be
insufficient for expected shock parameters.  However, they did not
discuss the possibility of external influences on the RSG wind, and
interpreted the drop in density as a shell resulting from a past
LBV-like eruption, as noted above. (They also noted that a change in
the degree of clumping might explain the drop in $N_{\rm H}$.)

An alternative explanation for the origin of a sudden jump in density
at a large radius may be an otherwise normal steady wind that becomes
confined by external pressure at large radii
\citep{gs96,chita08,mackey14}.  If the progenitor was indeed a more
massive star (above 18--20\,M$_{\odot}$) so that it was an O-type star
when on the main sequence, then its hot shocked main-sequence wind will
provide an external pressure that might decelerate the RSG wind and
confine it to a thin shell. (If the progenitor has similarly massive
O-type neighbours in a young stellar cluster, the hot interior of the
H~{\sc ii} region might create the required external pressure and
ionization.)  \citet{mackey14} have discussed the interesting
possibility that some SNe~IIn may arise from the interaction between
the SN shock and this type of externally confined, thin neutral shell.
In practice, this may be quite rare: most normal RSG progenitors are
at relatively low initial masses (i.e., 10--15\,M$_{\odot}$), which means
that they were not O-type stars on the main-sequence, and they reach
the RSG phase long after associated O-type stars have died.  On the
other hand, more-massive RSGs that die sooner have much higher wind
momentum in the RSG phase than assumed in the models, so it remains
unclear if normal RSG winds are likely to yield SNe~IIn through this
process.  

Although this scenario was proposed as a way to explain
SNe~IIn without appealing to eruptive pre-SN mass loss, it may also
provide a way for an interacting SN to be ``rejuvenated'' in its old
age, since the stalled shell is denser than a freely expanding wind.
\citet{mackey14} estimate that such a shell can have its density
enhanced by a factor of 80, and may contain a significant fraction
(perhaps a third) of the total mass lost during the RSG phase.  Their
models demonstrate that a SN can show a bolometric luminosity spike at
very late times from such a collision.  In this case it is difficult
to estimate the age of the shell, since its expansion may have
stalled, and its radius depends on the external pressure as much as on
the momentum of the RSG wind. Future monitoring of SN~2005ip may help
to test this hypothesis, as the shock continues to make its way
through this dense neutral shell and out into the ionized portion of
the RSG wind, or if SN~2005ip resumes a very slow decline as SN~1988Z
did.  Also, higher-resolution spectra may be able to place tighter
constraints on the pre-shock CSM expansion speed, since the narrow
components are unresolved in our moderate-resolution spectra presented
here.

Regardless of the true explanation for the density fluctuation in
SN~2005ip's outer CSM, it is useful to realize that mass-loss rates
this high (several times $10^{-4}$ to $10^{-3}$\,M$_{\odot}$\,yr$^{-1}$
or more) are quite extreme, and do not correspond to normal RSGs like
Betelgeuse, where the wind is two orders of magnitude less dense
\citep{smith09b}.  The cool evolved stars with winds this dense are
rare (see, e.g., \citealt{smith14} for an overview of mass-loss
rates), corresponding to the most luminous and extreme RSGs that are
enshrouded by their own dusty winds, like VY~CMa, NML~Cyg, S~Per,
etc., or to self-enshrouded super-asymptotic giant branch (AGB) stars.  
In the cases of SN~1988Z
and SN~2005ip, however, the very large CSM mass needed for their
sustained high-luminosity CSM interaction has been estimated as 
10\,M$_{\odot}$ or more.  This confidently rules out super-AGB stars
(total mass including a neutron star of only $\sim8$\,M$_{\odot}$) as
their possible progenitors.  This would be unlikely anyway, as these
super-AGB stars that may undergo O-Ne-Mg core collapse in an
electron-capture event are expected to produce a low SN kinetic energy
of $\sim10^{50}$ ergs \citep{nomoto82,nomoto87}.  Thus,
electron-capture SNe that produce SNe~IIn through interaction may be
expected to fade quickly (see, e.g., \citealt{mauerhan13}).  SN~1988Z
was hyperenergetic; up to day 3000, \citet{art99} estimate a total
radiated energy for SN~1988Z of $10^{52}$\,erg (model-dependent
$E_{\rm rad}$), and at least $2\times10^{51}$\,erg (observed $E_{\rm
  rad}$).  SN~2005ip appears very similar in many respects, and is
even more luminous and slower to decline in X-rays.

The extreme RSGs mentioned have initial masses of order 
25--35\,M$_{\odot}$, higher than normal unobscured RSGs, with more
tightly bound cores that may require more energetic explosions than
normal SNe~II-P.  This reinforces earlier suggestions that these
extreme RSGs (as opposed to normal RSGs) are the likely progenitors of
some SNe~IIn, including long-lived SN~1988Z-like events
\citep{smith09a,smith09b,vandyk+93,williams+02}.  More massive LBV
stars may also be able to supply the required amount of CSM mass
\citep{smith14,so06}.  However, the steady late-time decline over
decades in the case of SN~1988Z argues in favour of a freely expanding,
slow, and sustained dense RSG wind, rather than the sporadic eruptive
mass loss characteristic of LBVs.  For SN~2005ip, future observations
of its continuing evolution through the extended CSM may help us
choose more confidently between an extreme RSG --- perhaps exposed to
external ionizing radiation from its environment --- or an eruptive
progenitor.  In any case, the strong variability in H$\alpha$ around
day 3500 in SN~2005ip has not been seen so clearly before at such late
times in SNe~IIn, and suggests that similar late-time observations of
other objects can provide important constraints on the variability in
their pre-SN mass loss.  

Strong and variable mass loss of the type seen here occurring
1000\,yr before death presents a challenge for models of pre-SN
evolution.  Once we have a larger number of known SNe~IIn with such
enduring strong CSM interaction, it would be interesting to
investigate their statistical proximity to H~{\sc ii} regions as
compared to other populations of SNe.

\smallskip\smallskip\smallskip\smallskip
\noindent {\bf ACKNOWLEDGMENTS}
\smallskip
\footnotesize

We thank an anonymous referee for helpful suggestions.  Support was
provided by the National Science Foundation (NSF) through grants
AST-1210599 and AST-1312221 to the University of Arizona.  C.D.K.'s
research receives support from NASA through Contract Number 1255094
issued by JPL/Caltech.  W.F.\ was supported by NASA through an
Einstein Postdoctoral Fellowship. The supernova research of A.V.F.'s
group at U.C. Berkeley is supported by Gary \& Cynthia Bengier, the
Richard \& Rhoda Goldman Fund, the Christopher R. Redlich Fund, the
TABASGO Foundation, and NSF grant AST-1211916.

We thank the staffs at the MMT and Keck Observatories for their
assistance with the observations. Observations using Steward
Observatory facilities were obtained as part of the large observing
program AZTEC: Arizona Transient Exploration and
Characterization. Some of the observations reported in this paper were
obtained at the MMT Observatory, a joint facility of the University of
Arizona and the Smithsonian Institution.  This research was also based
in part on observations made with the LBT. The LBT is an international
collaboration among institutions in the United States, Italy, and
Germany. The LBT Corporation partners are the University of Arizona on
behalf of the Arizona university system; the Istituto Nazionale di
Astrofisica, Italy; the LBT Beteiligungsgesellschaft, Germany,
representing the Max-Planck Society, the Astrophysical Institute
Potsdam, and Heidelberg University; the Ohio State University and the
Research Corporation, on behalf of the University of Notre Dame,
University of Minnesota, and University of Virginia. Some of the data
presented herein were obtained at the W.M.  Keck Observatory, which is
operated as a scientific partnership among the California Institute of
Technology, the University of California, and NASA; the observatory
was made possible by the generous financial support of the W.M. Keck
Foundation. The authors wish to recognize and acknowledge the very
significant cultural role and reverence that the summit of Mauna Kea
has always had within the indigenous Hawaiian community. We are most
fortunate to have the opportunity to conduct observations from this
mountain.

% REFERENCES

\end{document}